\documentclass[times,9pt,twocolumn]{scrartcl} 
\usepackage{times}
\usepackage{color,setspace,epsfig,makecell}
\usepackage[numbers,sort&compress,round]{natbib}
\usepackage{graphicx}
\usepackage{amssymb,amsfonts,amsmath}
\usepackage{lineno}
\usepackage{caption}
\usepackage[numbers,sort&compress]{natbib}
\usepackage{booktabs} 
\usepackage{authblk}

\makeatletter
  \def\@eqnnum{{\normalfont \normalcolor [\theequation]}}
\makeatother

\newcommand{\eq}[1]{Eq.\,\eqref{#1}}
\newcommand{\eqs}[2]{Eqs.\,\eqref{#1}~\&~\eqref{#2}}
\newcommand{\eqss}[2]{Eqs.\,\eqref{#1}\,-\,\eqref{#2}}
\newcommand{\fig}[1]{Fig.\,\ref{#1}}

\begin{document}

\title{Stochastic evolutionary games in dynamic populations}

\author[1]{Weini Huang}
\author[2]{Christoph Hauert}
\author[1,*]{Arne Traulsen}
\affil[1]{Research Group for Evolutionary Theory, Max Planck Institute for Evolutionary Biology, August-Thienemann-Stra{\ss}e 2, 24306 Pl\"{o}n, Germany}
\affil[2]{Department of Mathematics, The University of British Columbia, 1984 Mathematics Road, Vancouver, B.C., Canada, V6T 1Z2}
\affil[*]{traulsen@evolbio.mpg.de}
\date{}
\maketitle

\section*{Abstract}
Frequency dependent selection and demographic fluctuations play important roles in evolutionary and ecological processes. 
Under frequency dependent selection, the average fitness of the population may increase or decrease based on interactions between individuals within the population.
This should be reflected in fluctuations of the population size even in constant environments. 
Here, we propose a stochastic model, which naturally combines these two evolutionary ingredients by assuming frequency dependent competition between different types in an individual-based model. 
In contrast to previous game theoretic models, the carrying capacity of the population and thus the population size 
is determined by pairwise competition of individuals mediated by evolutionary games and demographic stochasticity.
In the limit of infinite population size, the averaged stochastic dynamics is captured by the deterministic competitive Lotka-Volterra equations.
In small populations, demographic stochasticity may instead lead to the extinction of the entire population. 
As the population size is driven by the fitness in evolutionary games, 
a population of cooperators is less prone to go extinct than a population of defectors, 
whereas in the usual systems of fixed size, 
the population would thrive regardless of its average payoff. 

\section*{Introduction}
All natural populations are composed of a finite number of individuals. 
These individuals can reproduce, interact, die or migrate, which leads to changes in the population size over time. 
In many theoretical models, it is convenient and possible to neglect the effect of demographic fluctuations 
by assuming infinite populations when population sizes are sufficiently large \cite{taylor:MB:1978} 
or by assuming constant population size as in the Moran or Wright-Fisher process \cite{nowak:Nature:2004}. 
However, such simplifications may be invalid when considering additional ecological processes including 
oscillations in population size of predator and prey systems \cite{cohen:PRSB:1990,krebs:Science:1995,gilg:Science:2003}, 
periodic fluctuations and outbreaks of infectious diseases in humans \cite{rohani:Science:1999}, 
or chaotic dynamics under multi-species interactions \cite{beninca:Nature:2008}. 
The Lotka-Volterra equations provide a deterministic description of the abundances of species as continuous densities but they are not designed to include the impact of random drift.  
Theoretical models coupling changing population size and stochastic dynamics arising from individual based models have become more popular only recently \cite{Mao:JMAA:2003,mckane:PRE:2004,ovaskainen:TREE:2010,black:tree:2012,gokhale:BMCEvolbio:2013}.

The Lotka-Volterra equations naturally take frequency dependent selection into consideration. 
Under frequency dependent selection, the fitness of a given type (or species) depends on the composition of the entire population (or community) \cite{nowak:book:2006}. 
Different kinds of frequency dependence can lead to different dynamical patterns. 
Most prominently, negative frequency dependent selection can result in a stable coexistence of different types \cite{ayala:ARES:1974,gigord:PNAS:2001,nowak:Science:2004,koskella:Evolution:2009,huang:NatComm:2012}. 
One elegant way to describe such frequency dependence is through evolutionary game dynamics \cite{maynard-smith:book:1982,nowak:book:2006}. 
Evolutionary game theory has extensively developed the theory of stochastic dynamics in the past decade \cite{nowak:Nature:2004}. 
However, most progress has been accomplished for constant, finite population sizes, which is mathematically convenient, but not always reflects biologically appropriate scenarios. Game theoretic models that take changing population sizes into account mostly focus on deterministic dynamics \cite{hauert:PRSB:2006} -- similar to traditional ecological models. 
Here we introduce a simple and elegant model of stochastic evolutionary game dynamics that explicitly allows for changing population size through a natural interpretation of payoffs in terms of competition between individuals. 

Traditional game theoretic models assume that individuals obtain payoffs from interactions with other members in the population, which are translated into fitness. Individuals with higher fitness are assumed to have more offspring and hence reproduction is frequency dependent. Conversely, death rates are assumed to be constant or normalized such as too keep the population size constant. 
In contrast, here we focus on a microscopic description of the dynamics in terms of reaction kinetics equations. 
Assuming constant birth rates and frequency dependent death rates allows to interpret payoffs directly and naturally in terms of reaction rates, where selection acts on survival rather than reproduction. 
The present setup lends itself to a straight forward derivation of the deterministic dynamics in the limit of large population sizes in the form of competitive Lotka-Volterra equations, but equally allows to model the stochastic dynamics in finite populations of variable sizes, which may even lead to the extinction of the population.

\section*{Model and Results}

\subsection*{Stochastic dynamics}
Most models for stochastic evolutionary game dynamics consider a fixed population size, such that every birth is balanced by the death of another individual \cite{nowak:Nature:2004}.
Simply decoupling birth and death events in such models leads to random fluctuations in the population size and thus eventually to stochastic extinction \cite{gardiner:book:2004}. 
Instead, here we propose a framework based on the microscopic processes of birth, death and competition.
For simplicity, we focus on two types of individuals, $X$ and $Y$, but note that the generalization to arbitrary numbers is straight forward. Every individual reproduces
\begin{equation}
\label{reactionsBirthDeath5-1}
X\xrightarrow{}X+X 
\qquad \text{and}\qquad 
Y\xrightarrow{}Y+Y
\end{equation}
at constant rates $\lambda_{x\to xx}$ and $\lambda_{y\to yy}$, respectively, and dies
\begin{equation}
\label{reactionsBirthDeath5}
X\xrightarrow{}0
\qquad \text{and}\qquad 
Y\xrightarrow{}0
\end{equation}
also at constant rates $\lambda_{x\to 0}$ and $\lambda_{y\to 0}$, respectively. Competitive interactions result in four more processes
\begin{subequations}
\begin{eqnarray}
\label{reactionsBirthDeath5-2}
X+X \xrightarrow{}X
\qquad &\text{}&\qquad 
X+Y \xrightarrow{}Y\\
X+Y \xrightarrow{}X
\qquad &\text{}& \qquad 
Y+Y \xrightarrow{}Y.
\end{eqnarray}
\end{subequations}
In the simplest case, all competition rates are equal, such that two randomly chosen individuals compete for survival \cite{mckane:PRE:2004}.

\subsection*{Competitive selection}
The most natural way to introduce evolutionary games in the above framework is to relate the competition rates in \eq{reactionsBirthDeath5-2} to a payoff matrix
\begin{equation}
\label{eq:payoffs}
\bordermatrix{
  & \mathrm{X} & \mathrm{Y} \cr
\mathrm{X} & a & b \cr
\mathrm{Y} & c & d \cr},
\end{equation}
which determines the strength of competition between two $X$, $Y$ individuals as $a, b, c,$ and $d$ such that individuals obtaining higher payoffs are less likely to die in competitive interactions. More specifically, we assume that reaction rates are the inverse payoffs scaled by $M$, a unit for controlling population size,
\begin{eqnarray*}
&\lambda_{xx\to x}=\frac{1}{aM}, \qquad \lambda_{xy\to y}=\frac{1}{bM}, \\
&\lambda_{xy\to x}=\frac{1}{cM}, \qquad \lambda_{yy\to y}=\frac{1}{dM}.
\end{eqnarray*}
Thus, if, for example, an $X$ and a $Y$ individual compete, the $X$ dies with a probability proportional to $1/(bM)$ and the $Y$ proportional to $1/(cM)$. 
This requires that $a,b,c,d>0$ in order to remain meaningful in terms of reaction rates. 
In traditional models, payoffs are associated with reproduction, whereas here they refer to the probability of surviving competitive interactions. In both scenarios, high payoffs result in increased reproductive output over the lifetime of an individual. 
Naturally, reaction rates could follow a different functional dependence, for example $\lambda_{xx\to x}=\exp(-{aM})$, which would lead to qualitatively similar results as long as rates decrease with increasing payoffs, but without the restriction of positive payoffs.
The scaling term $M$ determines the frequency of competitive interactions as compared to birth or death events, \eqs{reactionsBirthDeath5-1}{reactionsBirthDeath5}. As long as population sizes are much smaller than $M$, competition is rare and most events are births or deaths. 
In large populations, however, competition becomes common and results in density dependent regulation of the population size. 

Implementing evolutionary games through competition is, of course, just one approach to link payoff matrices to reaction rates. 
Intuitively, it is tempting to assume that evolutionary games determine the birth rates instead because payoffs then more directly reflect fitness advantages. 
However, this requires microscopic reactions of the form $X+X\to X+X+X$, which occurs at rate $a$, etc. 
Such interactions seem more appealing in sexually reproducing populations rather than for the more traditional models, which focus on one sex only or are based on asexual reproduction. 
More importantly, however, even when including competition at fixed rates, this setup remains inherently biologically not meaningful, because it either results in extinction or indefinite growth of homogeneous populations (see Supporting Information, SI, Sec.\ 1). 
In natural systems, there can be positive feedbacks between different types \cite{bianchi:TREE:1989}, but they typically refer to different systems where other effects,
such as predation, guarantee that the population size remains finite. 
Here, we focus on a competitive system with negative feedback instead.

\subsection*{Large population size}
The reaction based system above can be formulated in terms of a continuous-time Master equation, see SI, Sec.~2. 
For large $M$, a measure of the equlibrium densities, this equation can be approximated by a Fokker-Planck equation, 
which describes the dynamics of the probability distribution of the system \cite{kampen:book:1997}. 
When the population densities approach the equilibria, we recover deterministic rate equations from the microscopic processes defined in \eqss{reactionsBirthDeath5-1}{reactionsBirthDeath5-2}
\begin{subequations}
\label{eq:NonlinearDEq2}
\begin{eqnarray}
\dot{x}&=& x \left(r_x-\frac{1}{a} \frac{x}{M}-\frac{1}{b}\frac{y}{M} \right)\\
\dot{y}&=& y \left(r_y-\frac{1}{c}\frac{x}{M}-\frac{1}{d}\frac{y}{M} \right),
\end{eqnarray}
\end{subequations}
where $x$ and $y$ denote the density of individuals of type $X$, $Y$ and 
$r_x=\lambda_{x \to xx}-\lambda_{x \to 0}$, $r_y=\lambda_{y \to yy}-\lambda_{y \to 0}$ indicate the intrinsic growth, i.e.\ the net growth rates from birth and death events, \eqs{reactionsBirthDeath5-1}{reactionsBirthDeath5}. 
Note that the deterministic limit can be derived directly based on the law of mass action.
Even though only the net growth rates, $r_x$ and $r_y$, enter \eq{eq:NonlinearDEq2}, 
it is important that the stochastic description does not lump the two processes together in either decreased net birth or net death rates. 
In particular, if spontaneous death events, \eq{reactionsBirthDeath5}, are dropped (or absorbed in reduced birth rates, \eq{reactionsBirthDeath5-1}) such that deaths occur only due to competition, \eq{reactionsBirthDeath5-2}, then populations would never go extinct in the stochastic formulation because the last individual standing would remain immortal.

In order to recover the familiar form of the competitive Lotka-Volterra dynamics \cite{zeeman:PAMS:1995,hofbauer:book:1998}, 
we factor out $r_x$ in \eq{eq:NonlinearDEq2}
and set $a=1$ (without loss of generality),
\begin{subequations}
\label{eq:NonlinearDEq3}
\begin{eqnarray}
\dot{x}&=& r_x\,x\left(1-\frac{x}{K}-\frac{1}{b}\frac{y}{K} \right)
\label{eq:NonlinearDEq3a}
\\
\dot{y}&=& r_y\,y\left(1-\frac{1}{c}\frac{r_x}{r_y}\frac{x}{K}-\frac{1}{d}\frac{r_x}{r_y}\frac{y}{K} \right).
\end{eqnarray}
\end{subequations}
Here $K=r_x M$ simply denotes the carrying capacity of $X$ types and $r_y d M$ the corresponding carrying capacity of $Y$ types. In the absence of $Y$ types \eq{eq:NonlinearDEq3a} reduces to the logistic equation
\begin{eqnarray}
\label{eq:logistic}
\dot{x}&=& r_x\,x\left(1-\frac{x}{K}\right),
\end{eqnarray}
which forms the basis for $r$-$K$-selection theory \cite{pianca:AmNat:1970},
where the carrying capacity $K$ is independent of the intrinsic growth rate, $r_x$. 
However, according to Eq.~\ref{eq:NonlinearDEq2}, $K$ is an emergent quantity determined by the population's environment \cite{kuno:RPE:1991,mallet:EER:2012, geritz:JMB:2012, novak:JTB:2013}, which crucially includes all members of the population together with their ecological interactions.
For example, if a mutant type $Y$ doubles its intrinsic rate of reproduction as compared to the resident $X$, $r_y=2 r_x$, then the mutant type readily displaces the resident and reaches its carrying capacity at twice the density of the resident, $K^y=r_y/r_x K=2K$, assuming that all other environmental parameters remain the same. 
This conclusion does not only follow from the microscopic description of relevant biological processes but has also been observed in experimental settings \cite{valle:RPE:1989,kuno:RPE:1991,bowers:EER:2003,mallet:EER:2012}. 
Moreover, in the following we show that the notion of a carrying capacity becomes even more challenging in populations of multiple types. 
Thus, we use the density of individuals at equilibrium instead of carrying capacity in heterogenous populations, where the total equilibrium density is $K_\text{cox}$ and the densities of type $X$ and type $Y$ individuals at this equilibrium are $K_\text{cox}^x$ and $K_\text{cox}^y$ respectively .

If $r_x<0$ or $r_y<0$, the corresponding type will invariably decline and disappear. Including competition only speeds up their demise. For example, this applies to the predators in the famous, oscillating Lotka-Volterra predator-prey dynamics.
Since negative interaction rates are not meaningful, at least three different types are required to observe oscillations in a competitive system \cite{hofbauer:book:1998,chiang:ITAC:1988}.

\vspace{-4mm}
\subsection*{Equilibria of the deterministic system} 

\begin{figure*}[tbhp]
\centering
 \includegraphics[width=1.0\textwidth]{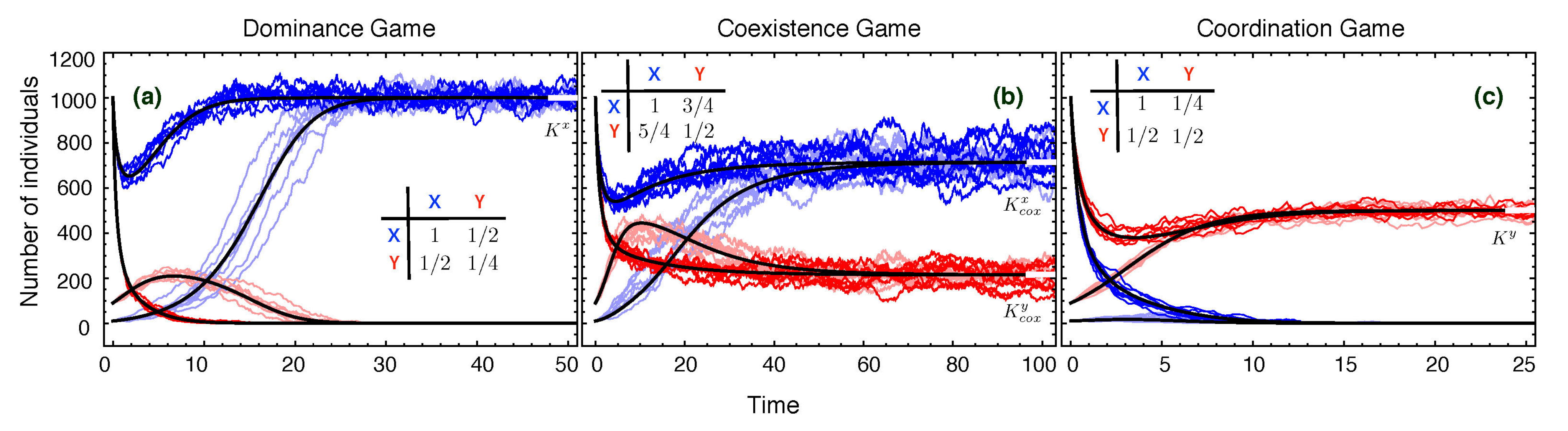}
\caption{
\label{FigDetStoc}
\sffamily
The stochastic dynamics fluctuates around the deterministic predictions in large populations (black lines). 
Each panel refers to one of the three characteristic classes of interactions as determined by the ranking of payoffs. Evolutionary trajectories for $X$ (blue) and $Y$ (red) types are shown for initially small ($x_0=10$, $y_0=90$, pale colours) and large ($x_0=1000$, $y_0=1000$ saturated colours) populations. 
Regardless of the game both types tend to increase for low initial population sizes, whereas both types decrease in densities at large initial sizes. In all cases the deterministic predictions agrees well with the stochastic dynamics in that the stochastic trajectories fluctuate around the deterministic average
(parameters $M=2000$, $\lambda_{x\to xx}=\lambda_{y\to yy}=0.6$, $\lambda_{x\to 0}=\lambda_{y\to 0}=0.1$, which translates into $K^x=1000$ and $K^y=250$ in \textbf{(a)} and $K^y=500$ in \textbf{(b)}, \textbf{(c)}).} 
\end{figure*}

The deterministic mean-field dynamics of our model serves as a valuable reference for the underlying stochastic evolutionary process. 
Birth and death rates, \eqs{reactionsBirthDeath5-1}{reactionsBirthDeath5}, may differ for different types but for the sake of simplicity and to highlight effects arising from evolutionary games, we discuss interactions of two types of individuals, $X$ and $Y$, with $r_x=r_y=r$. 
The evolutionary fate of each type depends on a combination of the strength of intra-type competition ($a$ and $d$) and inter-type competition ($b$ and $c$). 
In general, we can classify three different scenarios based on the payoff matrix: 
(i) For $a>c$ and $b>d$, type $X$ individuals invariably achieve higher payoffs (i.e.\ longer life expectancy) than $Y$ types and hence type $X$ dominates type $Y$. Similarly, type $Y$ dominates type $X$ whenever $a<c$ and $b<d$, see \fig{FigDetStoc}a. 
The prisoner's dilemma is the most prominent example of a dominance game \cite{maynard-smith:book:1982}. 
(ii) For $a>c$ and $b<d$ both types are at a disadvantage compared to the other type when rare. This reflects coordination games such as the stag-hunt game \cite{skyrms:book:2003}. 
(iii) Finally, for $a<c$ and $b>d$ both types have an advantage when rare but are at a disadvantage when abundant. 
Thus, an interior equilibrium exists where the two types co-exist, see \fig{FigDetStoc}b. 
The hawk-dove or snowdrift games are examples of such scenarios \cite{doebeli:EL:2005}. Note that even though the classification of the dynamics for two types is based on their payoffs in the same way as in the classical replicator dynamics (and the stability remains the same, see SI, Sec.\ 3), the position of the rest points in our deterministic system are naturally different. 
For example, in the replicator dynamics a coexistence game as in \fig{FigDetStoc}b exhibits a stable rest point at $x^\ast=(d-b)/(a-b-c+d) = 1/2$. 
In contrast, according to \eq{eq:NonlinearDEq3}, the frequency of $X$ at equilibrium is $x^\ast=10/13$. 
An intuitive reason for this increase in the relative abundance of $X$ is, that the total number of individuals at the mixed equilibrium is lower than the carrying capacity for a population of only $X$ types.

Therefore, the only possible equilibria are either homogenous $X$ or $Y$ populations or a stable heterogenous mixture of the two. 
According to \eq{eq:NonlinearDEq2} with $r_x=r_y=r$, the densities of individuals at the three equilibria are $K^{x}= aMr$ and $K^{y}= dMr$ as well as 
$K_\text{cox}=K^x_\text{cox}+K^y_\text{cox}$ with 
$K_\text{cox}^x=\frac{ac(b-d)}{bc-ad}Mr$ and 
$K_\text{cox}^y=\frac{bd(c-a)}{bc-ad}Mr$, which can be rewritten as $K_\text{cox}=K^x+\frac{(c-a)(b-a)}{bc-ad}K^y$. 
Note that in the co-existence equilibrium the density of individuals of each type is always lower than its carrying capacity in isolation, i.e. $K_\text{cox}^x<K^x$ and $K_\text{cox}^y<K^y$. However, the total number of individuals in mixed equilibria can either exceed or fall short of homogenous carrying capacities: if $b>a$, $K_\text{cox}>K^x$; if $b<a$, $K_\text{cox}<K^x$; if $c>d$, $K_\text{cox}>K^y$; if $c<d$, $K_\text{cox}<K^y$. More specifically, in co-existence games $b>a$ and $c>d$ holds such that the total number of individuals is highest in the mixed equilibrium, $K_\text{cox}>K^x,K^y$. 
Conversely, the reverse ranking is impossible: it would require $b<a$ and $c<d$ but this refers to coordination games where the mixed state is unstable and the population approaches either one of the homogenous configurations. 
Of course, in the corresponding stochastic realisations the population size at equilibrium is not fixed and instead fluctuates around the carrying capacity, see \fig{FigDetStoc}. 
For identical birth and death rates, the evolutionary game controls the relative growth or decline of the two types through competition, but regardless of the game the numbers of both types can increase or decrease if the current state of the population is far from equilibrium. All possible rankings of equilibrium densities are summarized in Table 1 in the SI. 

\subsection*{Stochastic simulations}
In contrast to the deterministic equilibrium predictions, the only evolutionary outcome in stochastic simulations is the eventual extinction of the entire population -- all other states are transient. 
Fortunately, the expected times to extinction rapidly 
grow with the density of individuals in equilibrium, controlled by $M$, \fig{EGD-Extinction-WholePop}. 
Hence, predictions based on deterministic dynamics, see \eq{eq:NonlinearDEq2}, keep providing valuable insights for the stochastic dynamics, see \eqss{reactionsBirthDeath5-1}{reactionsBirthDeath5-2}, especially for large population sizes. 
Substantial quantitative and even qualitative differences arise as illustrated \fig{FigDetStoc} for three characteristic types of interactions. The stochastic dynamics is implemented through the Gillespie algorithm \cite{gillespie:JCP:1976}. For the relatively large carrying capacity $K^x=1000$ in \fig{FigDetStoc}, each realisation of the stochastic dynamics fluctuates around the deterministic trajectory. Fluctuations represent an integral part of natural populations and hence stochastic evolutionary models provide a more natural way to study evolutionary trajectories, especially to capture the interplay between ecological and evolutionary processes \cite{parker:PRE:2009,ovaskainen:TREE:2010,black:tree:2012,eriksson:TPB:2013}.

In small populations, competition for survival is weak and the dynamics is mainly determined by the intrinsic growth rate, $r$, i.e.\ individual birth and death events. 
Consequently, small $r$ results in higher stochasticity but also tends to decrease the number of individuals at equilibra, which further amplifies the effect. 
As the population grows and approaches its carrying capacity, competition becomes increasingly important and competition rates, see payoff matrix \eq{eq:payoffs}, also control the size of fluctuations. 
Strong competition (small payoffs) reduces stochasticity, but also tends to decrease the population size, which may offset the reduction in terms of fluctuations.

Furthermore, if the numbers of the two types in the stochastic process are far away from the deterministic equilibrium, even the averaged stochastic dynamics can be very different from deterministic predictions. 
For example, in the deterministic case a dominant mutant  always succeeds in invading and eventually fixating in the population. 
In contrast, in the stochastic case a single mutant often fails to invade and fixate even if it is dominant. 
Note that fixation of a single mutant becomes even less likely in larger populations -- in spite of the fact that fluctuations decrease and the deterministic dynamics is recovered in the limit of large populations.

\subsection*{Extinction}
In ecological models the risk of extinction of a population due to demographic stochasticity has recently received considerable attention, see e.g. \cite{parker:PRE:2009,ovaskainen:TREE:2010,eriksson:TPB:2013}. 
Evolutionary game theory models demonstrated that stochastic fluctuations are important determinants for the fixation and extinction of individual traits even under constant population sizes \cite{nowak:Nature:2004} but remained unable to address the more dramatic possibility of the extinction of the entire population.

\begin{figure}[tbhp]
\centering
\includegraphics[width=0.5\textwidth]{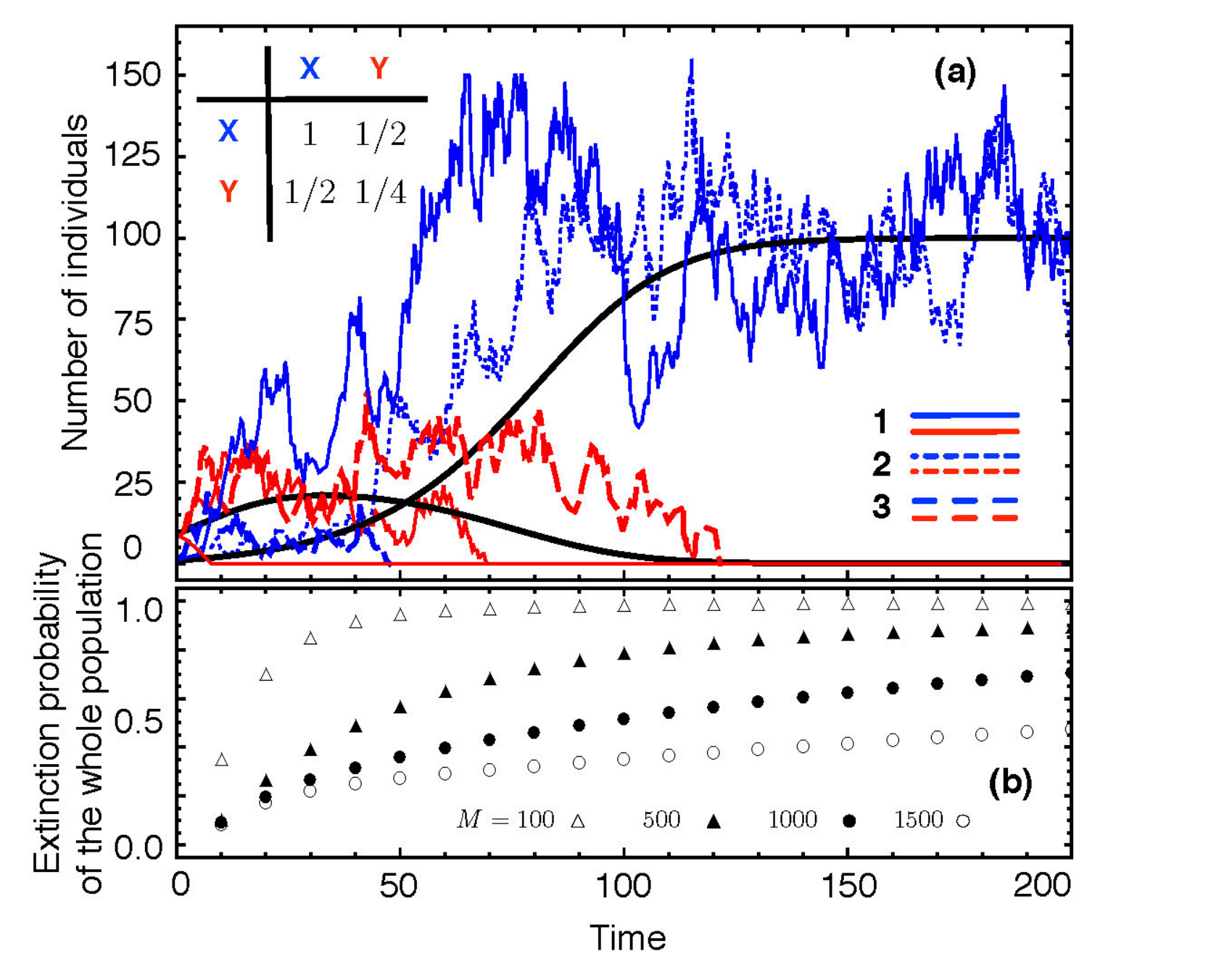}
\caption{
\label{EGD-Extinction-WholePop}
\sffamily
Stochastic dynamics of a dominance game in small populations. 
(a) Three realisations of the stochastic dynamics in a small population with two types, $X$ (blue) and $Y$ (red) for $M=1000$. Stochastic trajectories deviate significantly from the deterministic dynamics (solid black lines). For example, in the third realisation both types go extinct, first the dominant type $X$ followed type $Y$ a little later. 
(b) Cumulative probability for the extinction of the entire population over time, averaged over $10^5$ realisations for different $M$
(parameters $\lambda_{x\to xx}=\lambda_{y\to yy}=0.6$, $\lambda_{x\to 0}=\lambda_{y\to 0}=0.5$, $x_0=1$, $y_0=9$. This yields $K^x=100$ and $K^y=25$ in the upper panel).} 
\end{figure}

In a dominance game, the deterministic dynamics predicts that the dominant type invariably takes over the entire population and approaches its carrying capacity, see \fig{EGD-Extinction-WholePop}a. 
However, in stochastic models the two types $X$ and $Y$ may go extinct sequentially due to fluctuations. Especially when starting from small populations, the extinction probability is not negligible, see \fig{EGD-Extinction-WholePop}b. 
Note that the (cumulative) extinction probability converges to one as time goes on regardless of the population size. 
Only for larger populations extinction typically takes much longer and the extinction probability increases slowly with time. 
The smaller the population size -- due to small carrying capacities, fluctuations, or initial configurations -- the higher the risk of extinction.

In \fig{EGD-Extinction-WholePop}a, the dominant $X$ type has a higher carrying capacity than the $Y$ type and hence, provided that the $X$ mutant successfully invades and fixates, the population ends up more persistent. 
In the third realisation in \fig{EGD-Extinction-WholePop}a, the $X$ type is lost first and the entire population vanishes soon after. An essential factor for the persistence of small populations are the birth and death rates, \eqs{reactionsBirthDeath5-1}{reactionsBirthDeath5}, or, more specifically, their difference: for large intrinsic growth rates, small populations are more likely to escape extinction because the population can more readily recover and return to its carrying capacity. 

\begin{figure}[tbhp]
\centering
 \includegraphics[width=0.5\textwidth]{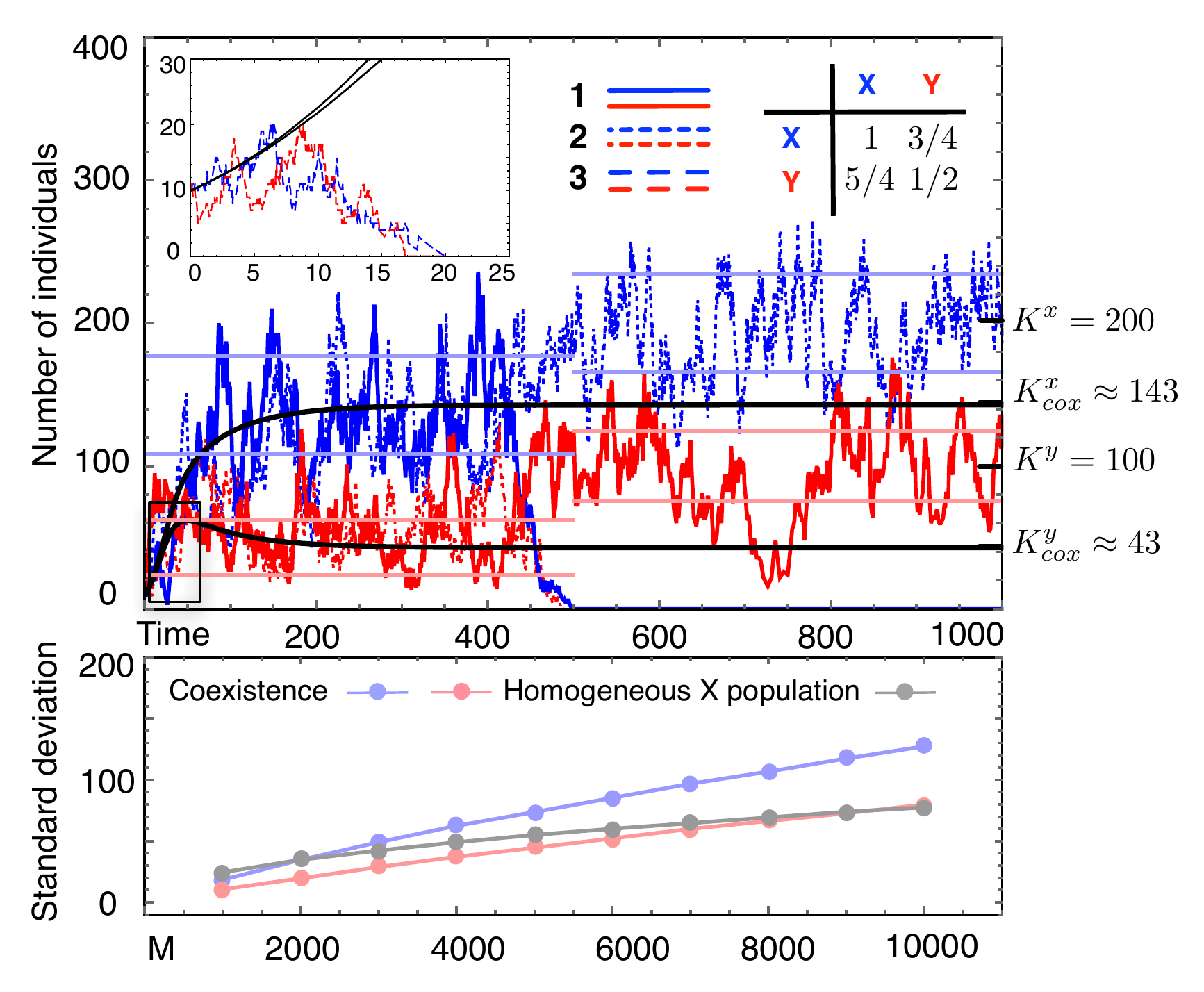}
 \vspace{-4mm}
\caption{
\sffamily
\label{EGD-Extinction-WholePop-CoexistenceGame}
Stochastic dynamics and large demographic fluctuations in a co-existence game. 
Top ($M=2000$): According to the deterministic dynamics (solid black lines) the two types $X$ (blue) and $Y$ (red) 
co-exist at their respective equilibrium, $K_\text{cox}^x$ and $K_\text{cox}^y$. 
The three stochastic realisations, however, reveal a different picture: 
In the first case, type $X$ goes extinct first, which benefits type $Y$ 
and results in a higher carrying capacity than the density of type $Y$ individuals in the co-existence equilibrium, $K^y>K^y_\text{cox}$. 
In the second case, type $Y$ goes extinct first and the new carrying capacity $K^x$
even exceeds the whole population size in the co-existence equilibrium $K^x>K_\text{cox}$. 
In the third case (inset), both types go extinct early on.
Pale lines represent the simulated standard deviation 
(averaged over $500$ independent simulations with $\geq 10^6$ data points each). 
Bottom: The standard deviation in the coexistence equilibrium increases with $M$, but 
differently for the two types
(parameters $\lambda_{x\to xx}=\lambda_{y\to yy}=0.6$, $\lambda_{x\to 0}=\lambda_{y\to 0}=0.5$, $x_0=10$, $y_0=10$).
}
\end{figure}

In co-existence games, the two types $X$ and $Y$ typically co-exist and their densities fluctuate around their respective (deterministic) densities of individuals in equilibrium, $K_\text{cox}^x$ and $K_\text{cox}^y$, see \fig{FigDetStoc}b. Whenever stochastic fluctuations drive one type to extinction, the carrying capacity of the remaining type changes to $K^x$ or $K^y$, respectively. Thus, the density of individuals of the entire population changes and is driven by the interplay of demographic fluctuations arising from intrinsic growth rates and the strength of competition, see \fig{EGD-Extinction-WholePop-CoexistenceGame}. 

In this case the extinction of the entire population is a two-step process: first one types goes extinct -- typically the type with the lower density of individual in equilibrium -- and then the population fluctuates around the homogenous carrying capacity of the remaining type such that the extinction dynamics is now governed by the stochastic equivalent for the logistic growth of a single type, which is well understood \cite{zeeman:PAMS:1995,parker:PRE:2009,ovaskainen:TREE:2010}.

\subsection*{Snowdrift game} 
As a concrete example of a co-existence game, we consider the stochastic dynamics of the snowdrift game \cite{doebeli:EL:2005} (or, equivalently, the hawk-dove game, \cite{maynard-smith:book:1982}). In the snowdrift game two individuals need to finish a task, which provides benefits $\beta$ to both. 
The costs of the task, $\gamma$, are shared equally if both cooperate, i.e. participate in completing the task. 
If only one participates, the cooperator has to bear the entire costs but the defector still receives the benefits. 
Finally, if both defect and refuse to participate, the payoffs for both individuals are close to zero. 
Hence, the payoffs of cooperators, $X$, and defectors, $Y$, are
$a=\beta-\gamma/2$, 
$b=\beta-\gamma$,
$c=\beta$, 
and 
$d=\varepsilon$,
where $\beta-\gamma\gg\varepsilon>0$ and $\beta>\gamma>0$. Note that in the limit $\varepsilon\to0$ the death rate of $Y$ types due to competition diverges and they are no longer able to persist in isolation. 
In order to maximize its own gain, each player should do the opposite of what its opponent does. 
A population of cooperators fluctuates around $K^x=(\beta-\gamma/2)r_x\,M$. 
A defecting mutant has a selective advantage and hence is likely to successfully invade and the population typically starts fluctuating around a new equilibrium, where the total density of individuals is lower-- from $b<a$ follows $K_\text{cox}<K^x$, see Table 1 in the SI. For sufficiently small $\varepsilon$, the conditions for the ranking $K^x>K_\text{cox}^x>K^y>K_\text{cox}^y$ are satisfied. Thus, even though defecting mutants are favoured, their abundances in the co-existence equilibrium is even lower than when in isolation. Consequently, stochastic fluctuations are more likely to eliminate defectors and re-establish cooperation at the original carrying capacity, $K^x$. For example, for $M=2000, r_x=r_y=0.5, \beta=1.5, \gamma=1, \varepsilon=0.05$, the density of cooperators $K_\text{cox}^x \approx 964$ in the mixed equilibrium is much larger than that of defectors $K_\text{cox}^y \approx 18$ and hence the odds of persistence are clearly in favour of cooperators.

\section*{Discussion}
Demographic fluctuations based on ecological interactions capture important aspects and represent crucial determinants of evolutionary trajectories, especially in smaller populations. Here, we introduce a simple stochastic framework built on the microscopic events of birth, death and competition. 
This framework admits a simple yet elegant way to implement evolutionary games through payoff based competition rates, which results in selection on survival instead of the more traditional fecundity based selection \cite{novak:JTB:2013}. 
This yields a stochastic model for evolutionary games in populations of changing and fluctuating finite size. 
In the limit of infinite population sizes, this framework recovers the deterministic dynamics of the competitive Lotka-Volterra equations and hence allows to pinpoint and emphasize differences that arise due to stochastic effects. 
The deterministic limit of the stochastic framework also highlights that, in contrast to the classical $r-K$-selection theory, the ecological carrying capacity of a population is an emergent quantity \cite{kuno:RPE:1991,mallet:EER:2012}, which depends on the population configuration and is determined by the underlying processes of birth, death and competition. In particular, mutations that alter the rates of any of these processes trigger a change in the (deterministic) carrying capacities of the mutant population, provided that it succeeds to take over, or of the mixed population in the case of co-existence. Our model implies that adaptation is not a simple process of accumulating beneficial mutations with higher carrying capacities in isolation, but instead an adaptive process that can favour invasion and fixation of mutations that are disadvantageous for the entire population, e.g. evolutionary suicide\cite{ferriere:PTRSB:2013}.

Dominant mutations are bound to take over with certainty under deterministic dynamics. However, in the stochastic scenario, the chances for a single beneficial mutant to successfully invade and take over remains small, even for a dominant strategy. At first it might be surprising that the chances of success decrease for increasing population sizes -- despite the fact that the limit of large populations recovers the deterministic dynamics. But, of course, in this limit the mutant density converges to zero, which resolves the apparent contradiction. 
Similar results can be found in classical models of finite populations with constant fitness values \cite{kimura:Genetics:1962}.
Here we investigated stochastic dynamics in well-mixed populations but a natural extension is to consider spatial dimensions, which may increase stochastic effects due to small local sub-populations \cite{durrett:TPB:1994,szabo:PR:2007,Mobilia:JSP:2007}.

Here, we have focussed on the paradigmatic case of one population and two types, but it is straight forward to extend the framework to include multiple types. 
In a population with three types, oscillations can persist in the stochastic process,
while the deterministic limit suggests cycles spiraling towards an internal equilibrium \cite{mckane:PRL:2005,Reichenbach:PRE:2006}, see Fig.1 in the SI. 
Moreover, our framework easily extends to group interactions, such as public goods games, by allowing for competitive interactions that involve more than two individuals. 
However, in either case, the number of microscopic interactions tends to increase rapidly and hence hampers a more general yet compact and intuitive presentation. 

The stochastic framework also emphasizes that in the long run populations invariably go extinct, which means that the deterministic equilibria merely indicate fleeting states -- albeit the expected time to extinction can be exceedingly long, especially for larger populations. Therefore, it remains reasonable to consider the deterministic predictions as a baseline superimposed by fluctuations of stochastic realisations. 
At the same time it is crucial, especially in smaller populations, to consider the persistence of individual traits or the viability of the entire population. 
For example, in the snowdrift game an invasion attempt by defectors triggers ecological feedback, which alters the carrying capacities in favour of cooperators, such that stochastic fluctuations help to eliminate defectors and re-establish homogenous cooperation.

\vspace{-3mm}
\subsection*{Acknowledgements}
C.H. acknowledges support from the Natural Sciences and Engineering Research Council of Canada (NSERC) as well as the Foundational Questions in Evolutionary Biology Fund (FQEB), grant RFP-12-10.
W.H. and A.T. acknowledge funding by the Max Planck Society.

\end{document}


\sffamily

\author{Weini Huang, Christoph Hauert, and  Arne Traulsen}
\date{}

\maketitle


\section{Relating payoffs to birth}

If payoffs affect birth instead of competition, the corresponding reactions can be written as 
\begin{eqnarray*}
X+X & \xrightarrow{a}X+X+X\\
X+Y & \xrightarrow{b}X+Y+X\\
X+Y & \xrightarrow{c} X+Y+Y\\ 
Y+Y & \xrightarrow{d}Y+Y+X.  
\end{eqnarray*}
Using the same principle as in the main text, i.e.\ individuals with higher payoffs have advantages in pairwise interactions, the birth rates should be an increasing function of the payoff 
elements $a,b,c$, and $d$.
The simplest choice is that birth rates equal payoffs. 
We assume constant intrinsic death rate $\lambda_d$ for both types, 
which implies 
\begin{equation*}
X\xrightarrow{\lambda_{d}}0 \quad {\rm and} \quad Y\xrightarrow{\lambda_{d}}0.
\end{equation*} 
Following a logistic growth model, we assume neutral competition
\begin{eqnarray*}
X+X & \xrightarrow{\lambda_c} X\\
X+Y & \xrightarrow{\lambda_c}X\\
X+Y & \xrightarrow{\lambda_c} Y\\ 
Y+Y & \xrightarrow{\lambda_c}Y.  
\end{eqnarray*}
Combining the ten reactions above, we obtain the deterministic rate equations 
\begin{eqnarray}
\label{eq:NonlinearDEq1}
\dot{x}&=& x \left(a\,x+b\,y-\lambda_c(x+y)-\lambda_{d}\right)\nonumber,\\\
\dot{y}&=& y \left(c\,x+d\,y-\lambda_c(x+y)-\lambda_{d}\right),
\end{eqnarray}
where $x$ and $y$ are the number of individuals of type $X$ and type $Y$. 
We denote the frequencies by 
$u=x/(x+y)$ and $v=y/(x+y)$. 
With Eqs.~\eqref{eq:NonlinearDEq1}, we obtain the change of the frequency of type $X$ as
\begin{eqnarray}
\label{eq:NonlinearDEq2}
\dot{u}&=&\frac{\dot x\,y -\dot y\,x}{(x+y)^2}\nonumber\\
&=& \underbrace{(x+y)}_{N}\underbrace{\,u(1-u)[(a-b-c+d)u+b-d]}_{\mathrm{Replicator \;\, dynamics}}.
\end{eqnarray}
The dynamics is equivalent to the replicator dynamics scaled by the total population size $N$. As $N$ changes over time, this can be considered as a dynamical, non-linear rescaling of time without changing the trajectories or equilibria from the standard replicator dynamics \cite{nowak:book:2006}. 

If we now focus on a homogenous population with type $X$ individuals only, i.e. $y=0$, in Eqs.~\eqref{eq:NonlinearDEq1}, we have 
\begin{equation}
\dot{x}= x \left((a-\lambda_c) x-\lambda_{d}\right). 
\end{equation}
This equation has two equilibria, $x=0$ and $x=\frac{\lambda_d}{a-\lambda_c}.$ 
The first equilibrium $x=0$, corresponding to extinction, is stable. 
The second equilibrium $x=\frac{\lambda_d}{a-\lambda_c}$ exists when $a>\lambda_c$.  However it is always unstable as $\dot x >0 $ for $x> \frac{\lambda_d}{a-\lambda_c}$ and $\dot x<0$ for $x<\frac{\lambda_d}{a-\lambda_c}$. 
In this alternative model where interactions affect birth, a homogenous population either goes extinct or explodes, depending on the initial population size. 
Although, the deterministic equations Eqs.~\eqref{eq:NonlinearDEq1} and \eqref{eq:NonlinearDEq2} appear to be reasonable, 
their ecological meaning remains unclear \cite{mallet:EER:2012}.


\section{Master equation and the diffusion approximation}

The microscopic process is a two dimensional Markov process in continuous time which can described by its master equation \cite{kampen:book:1997},
\begin{eqnarray}
\label{MasterEq1}
\frac{\partial P(x,y,t)}{\partial t}&=& T_{x-1}^+ P(x-1,y,t)+T_{y-1}^+P(x,y-1,t)\,\nonumber\\
&+&T_{x+1}^-P(x+1,y,t)+T_{y+1}^-P(x,y+1,t)\,\nonumber\\
&-&(T_{x}^++T_{y}^++T_{x}^-+T_{y}^-)P(x,y,t),\nonumber\\
\end{eqnarray}
where $P(x,y,t)$ is the probability that there are $x$ individuals of type $X$ and $y$ individuals of type $Y$ at time $t$, and $T$ is the transition rate of the population from one state to its neighbouring state. The subscript of $T$ refers to the type whose density changes, and the superscript denotes whether its density increases by one or decreases by one.
For example, $T_{x}^+$ is rate that the number of type $X$ increases from $x$ to $x+1$ and the number of type $Y$ remains constant and
$T_{y}^+$  is rate that the number of type $Y$ increases from $y$ to $y+1$ and the number of type $X$ remains constant.
The transition rates can be deduced from the reaction rates and the number of individuals of those types involved in the corresponding reactions,
\begin{eqnarray}
\label{TransRate1}
T_{x}^+&=&\lambda_{x\rightarrow xx}\,x\nonumber\\
T_{y}^+&=&\lambda_{y\rightarrow yy}\,y\nonumber\\
T_{x}^-&=&\lambda_{x\rightarrow 0}\,x+\frac{x^2}{aM}+\frac{x\,y}{bM}\nonumber\\
T_{y}^-&=&\lambda_{y\rightarrow 0}\,y+\frac{y^2}{dM}+\frac{x\,y}{cM}.
\end{eqnarray}
Here, $M$ is a scaling term which determines the frequency of competition compared to intrinsic growth. 
As we have shown in the main text, it controls the size of the system. 
When $M$ is larger, the density of individuals in the deterministic equilibria is larger. 

To perform a diffusion approximation of the master equation, we scale the numbers $x$ and $y$ by $M$, $\widetilde{x}=x/M$ and $\widetilde{y}=y/M$.
The new variables $\widetilde{x}$ and $\widetilde{y}$
 are approximately continuous for sufficiently large $M$. 
 We also rescale time as $\widetilde{t}=t/M$. 
This leads to
\begin{eqnarray}
\label{MasterEq2}
\frac{\partial P(\widetilde{x},\widetilde{y},\widetilde{t}\,)}{\partial \widetilde{t}}&=& T_{\widetilde{x}-\frac{1}{M}}^+  P(\widetilde{x}-\frac{1}{M},\widetilde{y},\widetilde{t}\,)+T_{\widetilde{y}-\frac{1}{M}}^+P(\widetilde{x},\widetilde{y}-\frac{1}{M},\widetilde{t}\,)\,\nonumber\\
&+&T_{\widetilde{x}+\frac{1}{M}}^-P(\widetilde{x}+\frac{1}{M},\widetilde{y},\widetilde{t}\,)+T_{\widetilde{y}+\frac{1}{M}}^- P(\widetilde{x},\widetilde{y}+\frac{1}{M},\widetilde{t}\,)\,\nonumber\\
&-&(T_{\widetilde{x}}^++T_{\widetilde{y}}^+ +T_{\widetilde{x}}^-+T_{\widetilde{y}}^- )P(\widetilde{x},\widetilde{y},\widetilde{t}\,),\nonumber\\
\end{eqnarray}
Note when we rescale the time, we also need to rescale the transition rates accordingly. 
Thus in the same time unit, the transition rate from, for example, from state $(x,y)$ to state $(x+1,y)$, equals to the transition rate from the scaled state $(\widetilde{x},\widetilde{y})$ and the scaled state $(\widetilde{x}+\frac{1}{M},\widetilde{y})$.
This leads to
\begin{eqnarray}
\label{TransRate2}
T_{\widetilde{x}}^+ &=&M\lambda_{x\rightarrow xx}\,\widetilde{x}\nonumber\\
T_{\widetilde{y}}^+ &=&M(\lambda_{y\rightarrow yy}\,\widetilde{y})\nonumber\\
T_{\widetilde{x}}^- &=&M(\lambda_{x\rightarrow 0}\,\widetilde{x}+\frac{\widetilde{x}\,^2}{aM}+\frac{\widetilde{x}\,y}{bM})\nonumber\\
T_{\widetilde{y}}^- &=&M(\lambda_{y\rightarrow 0}\,\widetilde{y}+\frac{\widetilde{y}\,^2}{dM}+\frac{\widetilde{x}\,\widetilde{y}}{cM}).
\end{eqnarray}

Now we expand the transition rates and probability densities in Eq. \eqref{MasterEq2} in a Taylor series at $\widetilde{x}$ or $\widetilde{y}$, and we obtain
\begin{eqnarray}
\label{FokPlankEq1}
P(\widetilde{x}\pm\frac{1}{M},\widetilde{y},\widetilde{t}\,)&\approx&P(\widetilde{x},\widetilde{y},\widetilde{t}\,)\pm\frac{\partial P(\widetilde{x},\widetilde{y},\widetilde{t}\,)}{\partial \widetilde{x}}\frac{1}{M}+\frac{\partial^2 P(\widetilde{x},\widetilde{y},\widetilde{t}\,)}{\partial \widetilde{x}^2}\frac{1}{2M^2}\,,\nonumber\\
P(\widetilde{x},\widetilde{y}\pm\frac{1}{M},\widetilde{t}\,)&\approx&P(\widetilde{x},\widetilde{y},\widetilde{t}\,)\pm\frac{\partial P(\widetilde{x},\widetilde{y},\widetilde{t}\,)}{\partial \widetilde{y}}\frac{1}{M}+\frac{\partial^2 P(\widetilde{x},\widetilde{y},\widetilde{t}\,)}{\partial \widetilde{y}^2}\frac{1}{2M^2}\,,\nonumber\\
T_{\widetilde{x}-\frac{1}{M}}^+ &\approx&T_{\widetilde{x}}^+-\frac{\partial\,T_{\widetilde{x}}^+}{\partial \widetilde{x}}\frac{1}{M}+\frac{\partial^2\,T_{\widetilde{x}}^+}{\partial \widetilde{x}^2}\frac{1}{2M^2}\,,\nonumber\\
T_{\widetilde{x}+\frac{1}{M}}^- &\approx&T_{\widetilde{x}}^-+\frac{\partial\,T_{\widetilde{x}}^-}{\partial \widetilde{x}}\frac{1}{M}+\frac{\partial^2\,T_{\widetilde{x}}^-}{\partial \widetilde{x}^2}\frac{1}{2M^2}\,,\nonumber\\
T_{\widetilde{y}-\frac{1}{M}}^+ &\approx&T_{\widetilde{y}}^+-\frac{\partial\,T_{\widetilde{y}}^+}{\partial \widetilde{y}}\frac{1}{M}+\frac{\partial^2\,T_{\widetilde{y}}^+}{\partial \widetilde{y}^2}\frac{1}{2M^2}\,,\nonumber\\
T_{\widetilde{y}+\frac{1}{M}}^- &\approx&T_{\widetilde{y}}^-+\frac{\partial\,T_{\widetilde{y}}^-}{\partial \widetilde{y}}\frac{1}{M}+\frac{\partial^2\,T_{\widetilde{y}}^-}{\partial \widetilde{y}^2}\frac{1}{2M^2}\,.
\end{eqnarray}
We denote 
$P=P(\widetilde{x},\widetilde{y},\widetilde{t}\,)$
and insert Eqs.~\eqref{FokPlankEq1} into Eq.~\eqref{MasterEq2}, then we obtain
\begin{eqnarray}
\label{FokPlankEq2}
\frac{\partial P}{\partial \widetilde{t}} &\approx&
\left(T_{\widetilde{x}}^+-\frac{\partial\,T_{\widetilde{x}}^+}{\partial \widetilde{x}}\frac{1}{M}+\frac{\partial^2\,T_{\widetilde{x}}^+}{\partial \widetilde{x}^2}\frac{1}{2M^2}\right)
\left(P-\frac{\partial P}{\partial \widetilde{x}}\frac{1}{M}+\frac{\partial^2 P}{\partial^2\widetilde{x}}\frac{1}{2M^2}\right)\nonumber\\
&+&\left(T_{\widetilde{y}}^+-\frac{\partial\,T_{\widetilde{y}}^+}{\partial \widetilde{y}}\frac{1}{M}+\frac{\partial^2\,T_{\widetilde{y}}^+}{\partial \widetilde{y}^2}\frac{1}{2M^2}\right)
\left(P-\frac{\partial P}{\partial \widetilde{y}}\frac{1}{M}+\frac{\partial^2 P}{\partial^2\widetilde{y}}\frac{1}{2M^2}\right)\nonumber\\
&+&\left(T_{\widetilde{x}}^-+\frac{\partial\,T_{\widetilde{x}}^-}{\partial \widetilde{x}}\frac{1}{M}+\frac{\partial^2\,T_{\widetilde{x}}^-}{\partial \widetilde{x}^2}\frac{1}{2M^2}\right)
\left(P+\frac{\partial P}{\partial \widetilde{x}}\frac{1}{M}+\frac{\partial^2 P}{\partial^2\widetilde{x}}\frac{1}{2M^2}\right)\nonumber\\
&+&\left(T_{\widetilde{y}}^-+\frac{\partial\,T_{\widetilde{y}}^-}{\partial \widetilde{y}}\frac{1}{M}+\frac{\partial^2\,T_{\widetilde{y}}^-}{\partial \widetilde{y}^2}\frac{1}{2M^2}\right)
\left(P+\frac{\partial P}{\partial \widetilde{y}}\frac{1}{M}+\frac{\partial^2 P}{\partial^2\widetilde{y}}\frac{1}{2M^2}\right)\nonumber\\
&-&\left(T_{\widetilde{x}}^++T_{\widetilde{y}}^++T_{\widetilde{x}}^-+T_{\widetilde{y}}^-\right)\,P
\end{eqnarray}
If we consider only the terms of the first order $M^{-1}$ and the second order $M^{-2}$ in Eq. \eqref{FokPlankEq2}, we obtain the Fokker-Planck equation
\begin{eqnarray}
\label{FokPlankEq3}
\frac{\partial P}{\partial \widetilde{t}}=&-&\frac{1}{M}\left(\frac{\partial }{\partial \widetilde{x}}\left(\left(T_{\widetilde{x}}^+-T_{\widetilde{x}}^-\right)P\right)+\frac{\partial }{\partial \widetilde{y}}\left(\left(T_{\widetilde{y}}^+-T_{\widetilde{y}}^-\right)P\right)\right)\nonumber\\
&+&\frac{1}{2M^2}\left(\frac{\partial^2}{\partial^2\widetilde{x}}\left(\left(T_{\widetilde{x}}^++T_{\widetilde{x}}^-\right)P\right)+\frac{\partial^2}{\partial^2\widetilde{y}}\left(\left(T_{\widetilde{y}}^++T_{\widetilde{y}}^-\right)P\right)\right)\,.\nonumber\\
\end{eqnarray}
From Eqs. \eqref{TransRate2}, 
we have 
$T_{\widetilde{x}}^+-T_{\widetilde{x}}^-=M((\lambda_{x\rightarrow xx}-\lambda_{x\rightarrow 0})\widetilde{x}-\frac{\widetilde{x}^2}{aM}-\frac{\widetilde{x}\widetilde{y}}{bM})$,
$T_{\widetilde{y}}^+-T_{\widetilde{y}}^-=M((\lambda_{y\rightarrow yy}-\lambda_{y\rightarrow 0})\widetilde{y}-\frac{\widetilde{y}^2}{dM}-\frac{\widetilde{x}\widetilde{y}}{cM})$, 
$T_{\widetilde{x}}^++T_{\widetilde{x}}^-=M((\lambda_{x\rightarrow xx}+\lambda_{x\rightarrow 0})\widetilde{x}+\frac{\widetilde{x}^2}{aM}+\frac{\widetilde{x}\widetilde{y}}{bM})$, 
and $T_{\widetilde{y}}^++T_{\widetilde{y}}^-=M((\lambda_{y\rightarrow yy}+\lambda_{y\rightarrow 0})\widetilde{y}+\frac{\widetilde{y}^2}{dM}+\frac{\widetilde{x}\widetilde{y}}{cM})$. Putting these into Eq. \eqref{FokPlankEq3}, we can rewrite the Fokker-Planck equation as
\begin{eqnarray}
\label{FokPlankEq4}
\frac{\partial P(\widetilde{x}, \widetilde{y},\widetilde{t})}{\partial \widetilde{t}}=&-&\frac{\partial }{\partial \widetilde{x}}\left((\lambda_{x\rightarrow xx}-\lambda_{x\rightarrow 0})\widetilde{x}-\frac{\widetilde{x}^2}{aM}-\frac{\widetilde{x}\widetilde{y}}{bM}\right)P(\widetilde{x}, \widetilde{y},\widetilde{t}\,)\nonumber\\
&-&\frac{\partial }{\partial \widetilde{y}}\left((\lambda_{y\rightarrow yy}-\lambda_{y\rightarrow 0})\widetilde{y}-\frac{\widetilde{y}^2}{dM}-\frac{\widetilde{x}\widetilde{y}}{cM}\right)P(\widetilde{x}, \widetilde{y},\widetilde{t}\,)\nonumber\\
&+&\frac{1}{2M}\frac{\partial^2}{\partial^2\widetilde{x}}\left((\lambda_{x\rightarrow xx}+\lambda_{x\rightarrow 0})\widetilde{x}+\frac{\widetilde{x}^2}{aM}+\frac{\widetilde{x}\widetilde{y}}{bM}\right)P(\widetilde{x}, \widetilde{y},\widetilde{t}\,)\nonumber\\
&+&\frac{1}{2M}\frac{\partial^2}{\partial^2\widetilde{y}}\left((\lambda_{y\rightarrow yy}+\lambda_{y\rightarrow 0})\widetilde{y}+\frac{\widetilde{y}^2}{dM}+\frac{\widetilde{x}\widetilde{y}}{cM}\right)P(\widetilde{x}, \widetilde{y},\widetilde{t}\,)\,.
\end{eqnarray}
The equivalent stochastic differential equations \cite{gardiner:book:2004}, which
can often be handled in an numerically more efficient way, are
\begin{eqnarray}
\label{SDE1}
\frac{\partial\widetilde{x}}{\partial\widetilde{t}}&=&(\lambda_{x\rightarrow xx}-\lambda_{x\rightarrow 0})\widetilde{x}-\frac{\widetilde{x}^2}{aM}-\frac{\widetilde{x}\widetilde{y}}{bM} + \sqrt{\frac{\widetilde{x}}{M}\left(\lambda_{x\rightarrow xx}+\lambda_{x\rightarrow 0}+\frac{\widetilde{x}}{aM}+\frac{\widetilde{y}}{bM}\right)}\,\xi\nonumber\,,\\
\frac{\partial\widetilde{y}}{\partial\widetilde{t}}&=&(\lambda_{y\rightarrow yy}-\lambda_{y\rightarrow 0})\widetilde{y}-\frac{\widetilde{y}^2}{dM}-\frac{\widetilde{x}\widetilde{y}}{cM} + \sqrt{\frac{\widetilde{y}}{M}\left(\lambda_{y\rightarrow yy}+\lambda_{y\rightarrow 0}+\frac{\widetilde{y}}{dM}+\frac{\widetilde{x}}{cM}\right)}\,\xi\,,
\end{eqnarray}
where $\xi$ is Gaussian white noise with mean $0$ and variance $1$.
Note that the noise term vanishes when the population is close to extinction, 
but increases approximately linearly in the population density.

\section{The stability of the equilibria}
\noindent The deterministic rate equations for two types are
\begin{subequations}
\label{eq:NonlinearDEq0}
\begin{eqnarray}
\dot{x}&=& x \left(r-\frac{1}{a}\frac{x}{M}-\frac{1}{b}\frac{y}{M} \right)\,\\
\dot{y}&=& y \left(r-\frac{1}{c}\frac{x}{M}-\frac{1}{d}\frac{y}{M} \right),
\end{eqnarray}
\end{subequations}
where $x$ and $y$ denote the numbers of individuals of type $X$ and $Y$, and $r$ refers to the same intrinsic growth rate for both types.
For $\dot{x}=0$ and $\dot{y}=0$, we have four equilibria, $E_1= (0,0)$, $E_2= (0,dMr)$, $E_3= (aMr,0)$, $E_4=(\frac{ac(b-d)}{bc-ad}Mr, \frac{bd(c-a)}{bc-ad}Mr)$.
In the following, we perform a linear stability analysis of the four equilibria.
The Jacobian matrix at the equilibrium $(x^*,y^*)$ is 
\begin{eqnarray}
J(x^*,y^*)=\begin{bmatrix}
r-\frac{2x^*}{aM}-\frac{y^*}{bM}&-\frac{x^*}{bM} \\
-\frac{y^*}{cM}&r-\frac{x^*}{cM}-\frac{2y^*}{dM}
\end{bmatrix}.
\end{eqnarray}
\begin{itemize}
\item[(i)] For $E_1= (0,0)$, 
$
J(0,0)=\begin{bmatrix}
r&0\\
0&r
\end{bmatrix}. 
$
Thus, for any $r>0$, the two eigenvalues are positive and this equilibrium is unstable.

\item[(ii)] For $E_2= (0,dMr)$, 
$
J(0,dMr)=\begin{bmatrix}
r-\frac{d}{b}r&0\\
-\frac{d}{c}r&-r
\end{bmatrix} 
$
with eigenvalues $-r$ and $-(d-b)r/b$. 
If $d>b$, the equilibrium is stable. Otherwise, it is unstable.

\item[(iii)] For $E_2= (aMr,0)$, 
$
J(aMr,0)=\begin{bmatrix}
-r&-\frac{a}{b}r\\
0&r-\frac{a}{c}r
\end{bmatrix}
$
with eigenvalues $-r$ and $-(a-c)r/c$. 
If $a>c$, the equilibrium is stable. Otherwise, it is unstable.

\item[(iv)] For $E_4=(\frac{ac(b-d)}{bc-ad}Mr, \frac{bd(c-a)}{bc-ad}Mr)$, the Jacobian matrix in this equilibrium is
\begin{eqnarray}
J_{E_4}&=&\begin{bmatrix}
r-\frac{2(b-d)c}{bc-ad}r-\frac{d(c-a)}{bc-ad}r&-\frac{ac(b-d)}{b(bc-ad)}r \\
-\frac{bd(c-a)}{c(bc-ad)}r&r-\frac{a(b-d)}{bc-ad}r-\frac{2b(c-a)}{bc-ad}r
\end{bmatrix}\nonumber\\
&=&r\begin{bmatrix}
-\frac{c(b-d)}{bc-ad}&-\frac{ac(b-d)}{b(bc-ad)} \\
-\frac{bd(c-a)}{c(bc-ad)}&-\frac{b(c-a)}{bc-ad}
\end{bmatrix}.
\end{eqnarray}
Thus the eigenvalues $\lambda$ can be obtained given 
\begin{equation}
\label{EquationABC}
\Big(\underbrace{-\frac{c(b-d)}{bc-ad}}_{A}-\lambda \Big)\Big(\underbrace{-\frac{b(c-a)}{bc-ad}}_{B}-\lambda \Big)-\underbrace{\frac{ad(b-d)(c-a)}{(bc-ad)^2}}_{C}=0.
\end{equation}
We can write Eq.\eqref{EquationABC} as $\lambda^2-(A+B)\lambda+AB-C=0$, thus $\lambda=\frac{A+B}{2}\pm\sqrt{\frac{(A-B)^2}{4}+C}$.  

This equilibrium exists only if the numbers of individuals of both types are positive, i.e.  $\frac{ac(b-d)}{bc-ad}>0$ and $\frac{bd(c-a)}{bc-ad}>0$. This yields two cases: $b>d\,\,\&\,\, c>a$ and $b<d\,\,\&\,\, c<a$, which both result in $C>0$ and $A+B<0$.

Thus, the first eigenvalue $\lambda_1=\frac{A+B}{2}-\sqrt{\frac{(A-B)^2}{4}+C}$, is always negative. The second eigenvalue $\lambda_2=\frac{A+B}{2}+\sqrt{\frac{(A-B)^2}{4}+C}$ is negative if $\frac{(A+B)^2}{4}>\frac{(A-B)^2}{4}+C$, which can be simplified to $AB-C>0$. From Eq. \eqref{EquationABC}, we have 
\begin{eqnarray}
AB-C&=&\frac{c(b-d)}{bc-ad}\frac{b(c-a)}{bc-ad}-\frac{ad(b-d)(c-a)}{(bc-ad)^2}\nonumber\\
&=&\frac{(bc-ad)(b-d)(c-a)}{(bc-ad)^2}\nonumber\\
&=&\frac{(b-d)(c-a)}{(bc-ad)}.
\end{eqnarray}
Thus,  if $b>d\,\,\&\,\, c>a$, then $AB-C>0$ and the equilibrium is stable; if $b<d\,\,\&\,\, c<a$, the equilibrium is unstable.
\end{itemize}

In general, although the equilibria in the deterministic limit are different from those in the replicator dynamics, their stability remains the same.

\section{Ranking of equilibria in a coexistence game}

In a coexistence game where $a<c$ and $b>d$, two types stably coexist with each other in the deterministic system.  
Here, we list all possible ranking rankings of the carrying capacities in two homogenous populations and the equilibrium densities in a heterogeneous population, see \tab{CapRanking}. One particularly interesting ranking is $K^x>K^y>K_\text{cox}^y>K_\text{cox}^x$: a homogenous population of type $X$ reaches higher numbers than type $Y$ but in the mixed equilibrium type $X$ is maintained at lower numbers than type $Y$. In stochastic processes this becomes particularly important because smaller carrying capacities result in a higher risk of extinction. For this particular ranking, a mutant $Y$ has an increased chance to take over the entire population because in the mixed equilibrium $X$ is outnumbered and hence is more likely to go extinct due to stochastic fluctuations. If this happens, the number of type $Y$ individuals will increase until it reaches $K^y$. Over the course of this invasion, the total number of individuals changes from $K^x$ to $K_\text{cox}$ and ends at $K^y$. 

In co-existence games where $b<a$ holds, the ranking is $K^y<K^x<K_\text{cox}$ and the total population size first increases and then decreases. In co-existence games where $b>a$ holds, the ranking is $K^y<K_\text{cox}<K^x$ and hence the total population size continuously declines to the carrying capacity of type $Y$ in isolation. 
This implies that evolution is not a simple process of accumulating beneficial mutations, which have higher carrying capacities in isolation, but instead the adaptive process can favour invasion and fixation of mutations that are disadvantageous for the entire population \cite{huang:NatComm:2012}. 
Similar evolutionary patterns are apparent in the prisoner's dilemma.
\begin{table*}[tbhp]
\centering
\begin{tabular}{ccc}
\toprule
Conditions & Ranking \\
\midrule
$a>d$ \,\,\,and\,\,\, $ad(c-d)<bc(a-d)$  & $K^x>K_{\text{cox}}^x>K^y>K_{\text{cox}}^y$\,\,\,\,\,\,\,\,\,\,\,\, \\
\,\,\,\,\,\,\,\,\,\,\,\,$a>d$ \,\,\,and\,\,\, $ad(c-d)>bc(a-d)$ and  & $K^x>K^y>K_{\text{cox}}^x>K_{\text{cox}}^y$\,\,\,\,\,\,\,\,\,\,\,\, \\
              \,\,\,\,\,\,\,\,\,\,\,\,\,\,\,\,\,\,\,\,\,\,\,\,\,\,\,\,\,\,\,\,$b>c$  \,\,\,or\,\,\, $ad(c-b)<bc(a-d)$ &\\
$a>d$ \,\,\,and\,\,\, $ad(c-b)>bc(a-d)$ & $K^x>K^y>K_{\text{cox}}^y>K_{\text{cox}}^x$\,\,\,\,\,\,\,\,\,\,\,\, \\
$a<d$ \,\,\,and\,\,\, $ad(b-a)<bc(d-a)$ & $K^y>K_{\text{cox}}^y>K^x>K_{\text{cox}}^x$\,\,\,\,\,\,\,\,\,\,\,\, \\
\,\,\,\,\,\,\,\,\,\,\,\,$a<d$ \,\,\,and\,\,\, $ad(b-a)>bc(d-a)$ and  & $K^y>K^x>K_{\text{cox}}^y>K_{\text{cox}}^x$\,\,\,\,\,\,\,\,\,\,\,\, \\
              \,\,\,\,\,\,\,\,\,\,\,\,\,\,\,\,\,\,\,\,\,\,\,\,\,\,\,\,\,\,\,\,$b<c$  \,\,\,or\,\,\, $ad(c-b)>bc(a-d)$ &\,\,\,\,\,\,\,\,\,\,\,\, \\
 \,\,$a<d$ \,\,\,and\,\,\, $ad(c-b)<bc(a-d))$ & $K^y>K^x>K_{\text{cox}}^x>K_{\text{cox}}^y$\,\,\,\,\,\,\,\,\,\,\,\, \\
\bottomrule
\end{tabular}
\caption{
In a coexistence game ($a<c$ and $d<b$), different conditions on the payoffs lead to 
certain rankings between the carrying capacities at homogenous and heterogenous equilibria. 
Here we list all possible rankings  assuming identical intrinsic growth rates $r_x=r_y$.%
\label{CapRanking}}
\end{table*}%

\section{Cyclic dynamics}
\begin{figure}[tbhp]
\centering
 \includegraphics[width=0.8\textwidth]{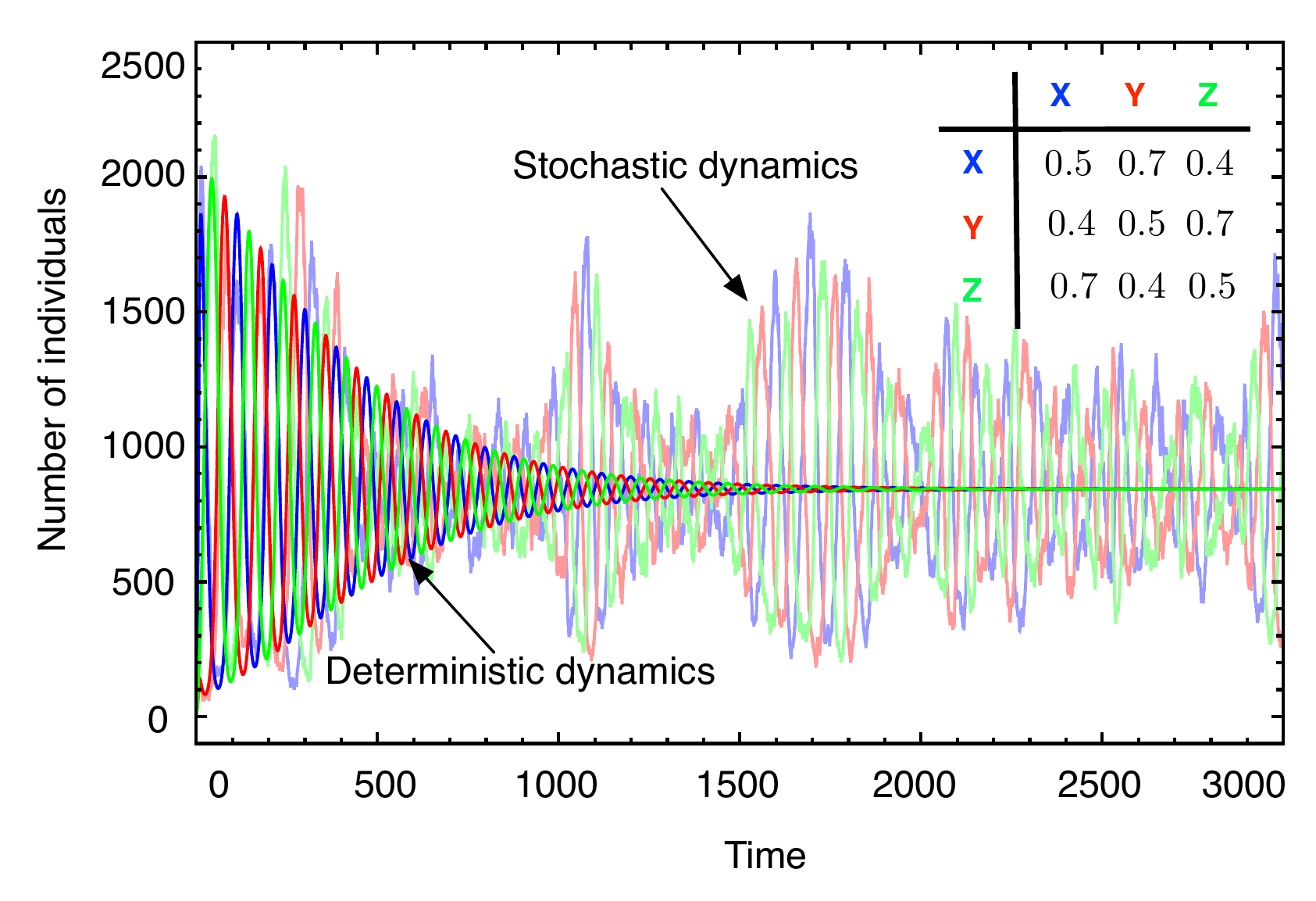}
\caption{
\label{CyclicDyn}
Cyclic dynamics of three types under competitive Lotka-Volterra dynamics. 
The interactions are given by the payoff matrix in the inset.
Saturated lines represent the deterministic dynamics, pale lines show one stochastic realisation
(parameters $\lambda_{x\rightarrow xx}=\lambda_{y\rightarrow yy}=\lambda_{z\rightarrow zz}=0.6$, $\lambda_{x\rightarrow 0}=\lambda_{y\rightarrow 0}=\lambda_{z\rightarrow 0}=0.1$, $M=10000$, $x_0=90$, $y_0=z_0=10$).
}
\end{figure}

In our stochastic model, cyclic oscillation can be observed in a population with three types. Here, we show an example where the cycles spiral into an internal equilibrium according to the deterministic equations, but periodic oscillations persist under the corresponding stochastic process, see \fig{CyclicDyn}. 

According to the payoff matrix (see inset in \fig{CyclicDyn}) in the deterministic dynamics, $X$ can invade a homogenous $Y$ population, $Y$ can invade a homogenous $Z$ population, and $Z$ can invade a homogenous $X$ population. The three types cycle into an internal equilibrium (see \fig{CyclicDyn}). 
However, demographic stochasticity drives the population away from the deterministic equilibrium and thus maintain the fluctuations over time. Note for small $M$,  the population size is so small that  demographic stochasticity will lead to the extinction of the whole population.

\section{Fixed population size}
Stochastic evolutionary dynamics under frequency dependent selection in populations of constant size, $N$, has recently attracted considerable interest, see e.g.~\cite{nowak:Nature:2004,traulsen:PRL:2005}. The most popular examples are based on birth-death models in discrete time. This results in a reduced set of microscopic interactions for two types,
$X+Y\xrightarrow{}X+X, X+Y\xrightarrow{}Y+Y$.
The deterministic rate equations are $\dot{x}=(\lambda_{xy \to xx}-\lambda_{xx \to xy})\,x\,y$ and $\dot{y}=-\dot{x}$.
In this case, $x$ and $y$ can be rescaled to indicate the frequencies due to constant population size. 
Constant reaction rates just lead to logistic growth of both types. Instead, to accommodate frequency dependent interactions, frequency dependent reaction rates are needed, e.g.  $\lambda_{xy\to xx}=f_x/N$ and $\lambda_{xy\to yy}=f_y/N$, where ${f_x}$ and ${f_y}$ represent the frequency dependent fitness. This yields $\dot{x}=x(1-x)(f_x-f_y)$, which is just the replicator dynamics \cite{hofbauer:book:1998}. 
Alternatively, setting $\lambda_{xy \to xx}=f_x/(\bar fN)$ and $\lambda_{xy \to yy}=f_y/(\bar fN)$ where $\bar f=x f_x+y f_y$ denotes the average fitness of the population, leads to $\dot{x}={x (1-x)(f_x-f_y)}/{\bar f}$ and recovers the adjusted replicator equation \cite{maynard-smith:book:1982}, which describes the deterministic limit of the frequency-dependent Moran process in discrete time \cite{traulsen:PRL:2005}. 
Interpreting frequency dependent interactions under fixed population size based on individual reactions is somewhat problematic. First, it is unintuitive to define frequency dependent reaction rates on the microscopic level because each reaction should occur independently. Second, for an evolving population we expect changes in the population size resulting by the changes of population composition rather than merely due to stochastic effects, which questions the basic tenet of models with constant population size.